          [2001/04/25 1.1 (PWD)]
\documentclass[a4paper,twocolumn]{esapub} 

\usepackage{times}
\usepackage{natbib}
\usepackage{amsmath}
\usepackage{epsfig}
\usepackage{txfonts}

\title{\huge{Asteroid and NEA detection models}}
\author{R.\ Michelsen}
\affil{Astronomical Observatory, NBIfAFG, Copenhagen University, Juliane Maries 
Vej 30, 2100 Copenhagen Denmark, E-mail: rene@astro.ku.dk}
\author{H.\ Haack}
\affil{Geological Museum, Copenhagen University, {\O}ster 
Voldgade 5-7, 1350 Copenhagen K, Denmark, E-mail: hh@savik.geomus.ku.dk}
\author{A.\ C.\,Andersen}
\affil{NORDITA, Blegdamsvej 17, 2100 Copenhagen, Denmark, E-mail: anja@nordita.dk}
\author{J.\ L.\ J{\o}rgensen}
\affil{{\O}rsted*DTU,MIS, Building 327, Technical University of Denmark,
2800 Lyngby, Denmark, E-mail: jlj@oersted.dtu.dk}

\begin{document}


\bibliographystyle{plain}

\maketitle

\begin{abstract}
We determine the possible detection rate of asteroids with the 
Bering mission. In particular we examine the outcome of the Bering 
mission in relation to the populations of Near-Earth Asteroids and 
main belt asteroids. This is done by constructing synthetic populations 
of asteroids, based on the current best estimates of the asteroid 
size-distributions. From the detailed information obtained from the 
simulations, the scientific feasibility of Bering is demonstrated and 
the key technical requirement for the scientific instruments on Bering 
is determined.
\end{abstract}

\section{Introduction}

The Bering mission is an autonomous mission, with the
purpose of making sample
observations of 
the inner asteroid populations.
In particular, Bering will travel through most of the space
between Venus and the outer parts of the asteroid belt at 3.5 AU
\cite{hansen2003},
and will thus be able to observe members of the Near-Earth Asteroid
(NEA) population, objects in the asteroid main belt, and objects 
\textit{en route}
from the main belt toward the NEAs.
The instruments on board the spacecrafts are the Advanced 
Stellar Compass (ASC) \cite{jorgensen2003a} that allows Bering to 
detect and 
follow moving objects, with the purpose of orbit determination
of the objects. In 
addition, the ASC can control a small telescope \cite{jorgensen2003b}, 
so that 
observations of the objects can be obtained. 
In this way,
Bering is able to provide both an orbit as well as a physical
characterisation of the objects. The key point is the autonomy
of the ASC and the telescope, that enables Bering to systematically
detect and
follow objects down to object diameters at the meter level
\cite{jorgensen2003a, jorgensen2003b}.
The scientific objectives have been discussed in detail elsewhere
\cite{andersen2003}.

One of the key issues to address is the two-fold
question of how many objects Bering will be able to observe.
First, this question is important to answer as a basis
for the scientific objectives, and with the need to describe the
scientific feasibility. Second, the number of objects detected
will depend on the limiting magnitude of the ASC, and is thus
vital for the technical requirements to the mission, also in terms
of the number of objects that needs to be processed by the 
autonomous spacecraft platform.
It is however not trivial to answer this question \cite{andersen2003},
hence we have a constructed a simulator capable of examining the
detailed aspects of the detections.

There already exists a number of simulations of the detection
of NEAs \cite{muinonen1998, jedicke2003}, however these simulations
differs from our needs in several ways. First of all, they are
made for ground based surveys for NEA discovery and follow-up.
For Bering, we need to be able to simulate observations made
by a spacecraft in an interplanetary orbit. 
Also, the simulations are normally strongly restricted in the size of 
the objects included.
For the Bering mission, we need to have detailed
knowledge of for how long time an object can be observed, as well
as the angular velocity. 
For instance, the trailing losses experienced
by ground based surveys due to fast moving objects across the field
of view during the exposure, are addressed by the Bering 
ASC capabilities to handle fast moving objects, however we need
to quantify the requirements to the ASC.

Thus, we in general need to understand how Bering will perform
when inserted into a given asteroid population, and with the possibility
of adjusting parameters, like the limiting detection magnitude
and the orbit.
We shall here focus on synthetic populations of 
the NEAs and the main belt asteroids. 
The synthetic objects
are treated as massless test particles in the simulation, and after an
initial sorting, the objects are numerically integrated. 
This allows a careful
examination of the requirements to the ASC when probing members
of these populations down to the meter-range.

\section{Asteroid magnitudes and diameters}

As discussed \cite{andersen2003}, one of the challenges when
observing asteroids are their rapid and drastic variations in
magnitude. The magnitude of an asteroid does not only depend
on physical parameters like the size and the albedo, but also
on the distances to the Sun and the observer, and the phase angle. 
Depending on the geometry, there can thus be a strong time dependency
on the magnitude variations.

The magnitudes of asteroids are normally described by a two-parameter
model. The absolute magnitude $H$ of an asteroid is defined as the
V-magnitude at unit distance to the Sun ($r$)  and the Earth ($\Delta$), 
and at
phase angle $\alpha = 0$. The second parameter is the slope parameter
$G$, which is an expression for the geometric albedo. Thus
the relation between the observed V-magnitude and $H$ and $G$ is
\cite{bowell1989}
\begin{equation}
\begin{split}
V =& H + 5 \log r \Delta \\
&- 2.5 \log[(1-G) \Phi_1 (\alpha) + G\Phi_2 (\alpha)]
\end{split}
\end{equation}
where $\Phi_1$ and $\Phi_2$ are functions of the phase angle $\alpha$.
$r$ and $\Delta$ are measured in AU.

Only for a few objects has the 
slope parameter $G$ and the
albedo $p_V$ 
been measured. The desire however persists to translate the
absolute magnitude $H$ into a diameter $D$ of the object. This can
be done by an \textit{a priori} assumption on the albedo of the object.
The diameter can then be expressed as \cite{brown2002,werner2002}
\begin{equation}
D({\rm km}) = 1329 \cdot 10^{-H/5} / \sqrt{p_V}
\end{equation}
where $D$ is in units of km.
Care is needed when evaluating the sizes of the smallest sub-kilometer
asteroids, as the actual albedo of these objects is currently not
known.

\section{The NEA population}

During recent years, many publications have been made in the
attempt of establishing the size distribution of Near-Earth Asteroids.
Bottke and collaborators \cite{bottke2000} have constructed a 
de-biased estimate
of the NEA size distribution, for the range up to absolute magnitude
$H=22$, by simulating a ground based NEA survey, and taking into 
account the nature of the source regions of the NEAs. 
For this range,
they find a best fit of the form $\log N = \alpha H + \beta$, where
$N$ is the accumulated number of objects and $\alpha$ and $\beta$
are constants. Werner \cite{werner2002} 
provides an estimate for the distribution up to $H=30$ ($D=3$m), 
derived
from the lunar crater size-frequency distribution. Their distribution
corresponds well to the estimates obtained from ground based NEO
surveys \cite{rabinowitz2000}. Werner obtains a distribution
similar to that of Bottke, however at around $H=21$ the 
distribution shows a turning point, with a much steeper slope.
For the even smaller end, Brown and collaborators \cite{brown2002} 
reports
satellite observations of bolide detonations in the Earth's atmosphere,
for objects in the range of 1m to 100m. The distribution derived
from these observations is
consistent with the work of Werner.

It should 
be stressed that the estimated populations are based on indirect
methods, and that various factors may influence the actual, present
size distribution. 
One of the factors is the translation from
$H$ to $D$, as the albedo $p_V$ of the NEA population is more
or less unknown.
There are indications, for instance
in the work of Werner, that the smallest NEAs have
larger albedo. This is compatible with our hypothesis that
the smallest asteroids are young collisional fragments without
surface dust, and with minimum exposure to cosmic radiation,
hence the objects have larger albedo.

\begin{figure}
   \centering
   \epsfig{file=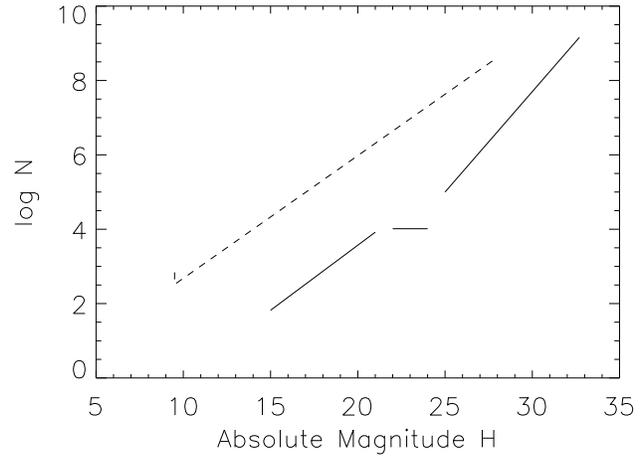,width=1.0\linewidth}
      \caption{The absolute magnitude ($H$) distribution of the asteroid populations
    of our simulations. N is the accumulated number of objects.
     The solid line is the NEA population, the dashed line is the 
main belt distribution. For the NEA population, we adopt a three-step
distribution, based on the indications of a steeper increase in the
number of the smaller objects, compared to the larger objects.
The main belt follows a simple linear distribution in ($H, \log N$),
however with certain limitations of this distribution (see text
for discussion).
              }
         \label{Fig:distrplot}
   \end{figure}

The distribution we have adopted for the NEA population is a
combination of the estimation by Bottke, combined with the
work by Werner, supported by the measurements by Brown.
We shall thus split the distribution into three parts, with bins
of 1 magnitude, as
\begin{subequations}
\begin{align}
\log N &= 0.35 H - 3.43    & ; \ \  & 15 \ge H < 22 \label{bottke}\\
N &= 10303    & ; \ \  & H=[22,24] \label{plateau} \\
\log N &= 0.54 H -8.5 & ; \ \ & 24 < H \leq 32.7 \label{werner}
\end{align}
\label{neapopulation}
\end{subequations}
The distribution can be seen in Fig.\ \ref{Fig:distrplot}.
The first part is thus adopted from Bottke, followed by
a plateau. The last part of the distribution is taken from the 
low-albedo distribution of Werner.
The $H=32.7$ has been chosen, so that with an albedo of $p_V = 0.17$,
this corresponds to a diameter of $D=1$m. The total number of 
objects in the distribution amounts to $\log N = 9.158$, or
approximately $1.4\times 10^9$ objects. For the
slope parameter we initially assumed $G=0.15$, a low-albedo
population, but later we made tests with a high-albedo population
using $G=0.32$, corresponding to $p_V \approx 0.25$ (see the
discussion of the results below).
 
Besides the distribution in terms of $H$, the de-biased distribution 
by Bottke also suggests a spatial distribution of
the objects, in terms of the eccentricity, the semi-major axis
and the inclination. However, for the above synthetic population,
we have selected a uniform distribution of the orbital elements
(within the definitions of a Near-Earth Object). In principle,
this is a failure of the distribution, relative to the current
best estimates. However, the suggested distribution of the 
orbital elements
is only valid for objects with $H<22$. There is only vanishing
statistical material for smaller objects (larger $H$), hence
there is no basis to assume a continuation of the distribution
in the orbital elements for $H>22$. In any case, the number
of objects with $H\leq 22$ amounts to approx.\ 18600, a vanishing
number compared to the total number of objects.

\section{The main belt population}

The main belt population is very much different from that of the NEAs. 
First of all, our observational knowledge is strongly restricted
to only the very largest objects, due to the far distances to the
objects. 

Durda and collaborators \cite{durda1998} have made numerical models of the
collisional evolution of the main belt asteroids. The observed
distribution shows two bumps relative to a linear $(D, \log N)$ fit,
for objects with $D=100$km and $D=5$km. By adjusting the parameters
of their models, Durda is able to reproduce these bumps,
and thus to obtain an estimate of the main belt distribution down to
1m in size. Other estimates for the distribution exists
(see \cite{davis2002} for an overview), and they
tend to agree for objects down to 1km. 

For these simulations we shall adopt the distribution by Durda,
with some modifications.
The focus of our simulations are on the sub-kilometer
objects, hence 
for the range $H=[5;10]$ we count the total number of known
objects \cite{jedicke2002}, and assign
to them the absolute magnitude $H=9.5$. This involves 694 of the
largest known objects, and causes an initial bump in the
distribution (Fig.\ \ref{Fig:distrplot}).
We then make a linear fit to the distribution of Durda 
for their size range 100m to 50km,
\begin{equation}
\log N = 0.33 H - 0.62
\end{equation}
and apply this distribution for the magnitude interval $H=[10;27.94]$ 
in steps of 0.5 magnitudes. Using the albedo referred to
by Durda of $p_V = 0.1178$, the $H=27.94$ corresponds to 
an object size of $D=10$m. However, for the slope 
parameter we
use $G=0.15$ $(p_V=0.17)$, a slightly larger albedo than 
used by Durda, so that the smallest object in our distribution 
in fact has $D=8$m.

This distribution has two flaws. First of all, it does not take
the mentioned known bumps of the large objects into account. 
However, the number of objects
among the large sizes neglected in this manner is completely negligible
in comparison with the total number of objects
in this synthetic population. The large objects are as well not
the main objective of this work, but should under all
circumstances be treated separately,
taking into account the already known objects.
The works by Durda and by Davis \cite{davis2002} seems to 
indicate an {\textit{increase}} in the slope of the size distribution
for $D<100$m. Whether this holds true or not is one of the
scientific objectives of Bering. However, it means that relative
to these estimates, our synthetic main belt population is heavily
underestimating the number of 10m objects, with a factor of 10 or
more. This must be taken into account, when reviewing the results
of the simulations. 

The total number of objects in the main belt population is thus
$\log N = 8.6$, or around $4\times 10^8$ objects. The orbital
elements are assumed to be uniformly distributed.

\section{The numerical approach}
\begin{figure}
   \centering
   \epsfig{file=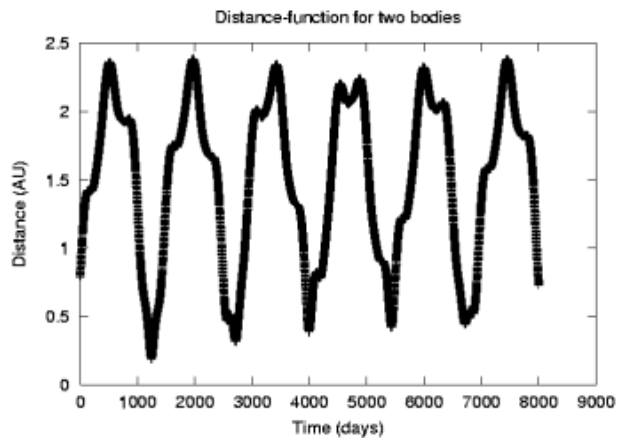,width=1.0\linewidth}
      \caption{Example of the distance variation between
               Bering and a synthetic object (a test particle). The 
              MOID is the 
               global minimum of the distance.
              }
         \label{Fig:exdist}
   \end{figure}

With the large number of objects involved in estimating
the detection rate, a direct numerical integration is not feasible
due to the very heavy demands on CPU time and disk space.
Instead, we adopted a solution that provide a 
rough sorting of the objects, discarding those that would not become
observable from Bering within a reasonable time. A simple idea 
is to calculate the minimum orbital intersection distance
(MOID) between the Bering orbit and every object. If the MOID
distance is sufficiently small, so that in a best case situation
the object is observable within the magnitude limits, the
object is selected for subsequent detailed analysis, 
otherwise it is discarded. Instead of a purely numerical
approach for determining the MOID, for instance by integrating
the orbits and calculating the smallest distance between 
Bering and the
objects, we use a semi-analytic approach, implementing the 
MOID method suggested by Sitarsky \cite{sitarsky1968}. 
For two Keplarian orbits, the distance function can be written as
\begin{equation}
f(V,v) = \tfrac{1}{2} (\mathbf{R} - \mathbf{r}) ^2,
\end{equation}
where $(V,v)$ are the true anomalies of the objects of concern,
and $\mathbf{R}, \mathbf{r}$ are the cartesian coordinates relative
to some origin. By examining the derivatives of this function,
Sitarsky derives analytical equations for the minimum
of the function, the MOID. These equations can, for the general case,
not be solved analytically, hence we apply a 
version of the Newton-Raphson method that is suited for non-linear
cases (Press et al.\ \cite{press1992}), and in particular
ensures convergence toward some minimum. This problem is however
not trivial. Often, the solution must be sought in very rough
terrain, Fig.\ \ref{Fig:exdist},
hence one set of initial conditions may not be enough to ensure
that the global minimum is found. Thus we solve the 
equations for several sets of initial conditions, and 
from the resulting local minima, 
we accept the smallest value as the MOID.
We verified the implementation of the MOID method against
a number of different orbit scenarios, 
and compared 
with
the minimum distance between the two
orbits, obtained by direct numerical 
integration.

After the MOID has been calculated, we apply 
a selection criteria for the sorting.
We calculate the $V$-magnitude of the asteroid by using  
the MOID as the distance to the observer.
The perihelion distance, the smallest distance to the
Sun, is used as the asteroid-Sun distance. Further, the phase
angle is set to zero. This is truly a best case configuration
of the asteroid. If the calculated magnitude is $V\leq V_{\rm lim}$,
where $V_{\rm lim}$ is the Bering detection limit, the object is 
saved, otherwise it is discarded.

This MOID-sorting is done for all the objects. 
It is computationally heavy, but allows focus on only
the most interesting objects. As the sorting can be quite
aggressive, depending on $V_{\rm lim}$, it can drastically decrease
the demands for computing time and disk space for a detailed
analysis of the Bering detection rate.

The objects that were selected by the MOID sorting for
further analysis, are
endured to a numerical integration. For each
time step, the V-magnitude of the objects is calculated, and compared
against $V_{\rm lim}$, disregarding the direction to the object,
thus assuming that Bering is able to monitor the whole sky. 
This is a good approximation, as the only directions Bering will
not cover are toward the Sun, in which direction only the
night side of the objects will be seen, thus being too faint
under all circumstances.
During the integration, book keeping is made
of which objects had $V<V_{\rm lim}$, the smallest $V$ of the objects
and the corresponding
tangential velocity during the minimum $V$. 
By running a small filter after the integration,
a complete list of the observable objects could be generated from
the book keeping data.

As integrator, we used the SWIFT package 
\cite{levison2003,levison1994}.
The
package was modified to allow handling of the magnitude calculations
and of the book keeping data. As the primary integrator we used the
symplectic RMVS3 method, however for selected objects we made a 
refined check with the classic Bulirsch-Stoer
method \cite{press1992}, without any noticeable difference in the
results of the two integrators.

Due to the very transient variations in brightness of the smallest
objects, the integrations were typically made over 50 days, with
a data dump and magnitude calculation every 0.02 days (28.8 minutes).

The MOID implementation and the numerical integrations were parallelized 
at the run-level, and ran 
on the IBM RS-6000 cluster of the Center for Scientific
Computing Aarhus (CSC-AA) \cite{cscaa2003}.

\section{Results for the NEAs}

The first key problem to deal with was to assess the
value of $V_{\rm lim}$. This value is critical for
the mission feasibility in terms of the 
the detection rate of the Bering Advanced Stellar Compass, 
i.e.\ the number
of objects that would be observed per unit of time. If $V_{\rm lim}$
is too small, only a small number of objects would be found
during the mission life time, and the
science telescope would remain idle most of the time.
If $V_{\rm lim}$ is too large, it would
exhaust the onboard system, for instance requiring faster CPUs
with larger power consumption, violating the power budget.

Due to the heavy computations involved of the full asteroid
populations,
we decided to initially work on a small subset of the NEAs,
by extrapolating the Bottke distribution, Eq.\ \eqref{bottke},
to an object size of around 1m, $H\approx 32.5$, for a total
of exactly $10^8$ objects. Many of the ground based
surveys are able to reach $V=19$, so this value we used as the
initial
$V_{\rm lim}$ in the 
MOID and the integration. As the orbit of Bering, we simply
used the Earth orbit, for comparison with the ground based surveys.
The results are shown as Run 1 in Table \ref{Tab:results}.
   \begin{table*}[]
     \centering
      \caption[]{Results of the simulations for the NEA and main
belt populations. $N$ is the initial number of objects, $N_{\rm MOID}$
is the number of objects remaining after the MOID sorting and
$N_{\rm obs}$ is the number of objects that could be seen during
the integration time for the stated values of $V_{\rm lim}$ and $G$.
$t$ is the time span and $t_{\rm dump}$ is the 
time resolution of the integration. For Run 3, the object
population was adopted from Run 2, however with a change in $G$.
The population of Run 5 is the same as for Run 4, however with
different parameters for the numerical integration.
}
         \label{Tab:results}
    \begin{tabular}{| l | c | c | c || c | c |}
    \hline
    & \multicolumn{3}{c||}{NEA population}  & \multicolumn{2}{c|}{Main belt population} \\
    \cline{2-6}
      & Run 1 & Run 2 & Run 3 & Run 4 & Run 5 \\
    \hline
    $V_{\rm lim}$ & 19.0 & 15.0 & 15.0 & 15.0 & 15.0 \\
            G & 0.15 & 0.15 & 0.32 & 0.15 & 0.15   \\ 
    $t$ (days) & 50 & 50 & 50 & 356 & 50 \\
    $ t_{\rm dump}$ (days) & 0.02 & 0.02 & 0.02 & 1.0 & 0.02 \\
    $N$ & $10^8$ & $\approx 1.4\times 10^9$ & -  &
                   $\approx 4\times 10^8$  & -  \\
    $N_{\rm MOID}$ & $\approx 7\times 10^6$ & $\approx 10^7$ & - &
                 $\approx 10^7$ & - \\
    $N_{\rm obs}$ & 619 & 16 & 23 & 647 & 352 \\
    \hline
    \end{tabular}
   \end{table*}
Of the initial 100 million objects, only 7 million, or 7\%,
remains after the MOID sorting. Of these objects, 619 were seen
by Bering over 50 days. 
The detection rate would thus be 12.4 objects/day. For each object
we know the circumstances of the detection, for instance
the smallest object observed had 
$H=32.66, D=1$m, and the object had $V<V_{\rm lim}$ for 144 minutes.
At the brightest point $V=18.5$, and the object was moving with
25.2~"/sec. 
For each of the 619 objects we thus know the maximum brightness 
of the objects during the 50 days, and we can then determine
the threshold of $V$ where the detection rate goes to zero.
In this manner, we find that for $V_{\rm lim}=15$, 17 objects
could be observed, corresponding to a detection rate of 0.3 objects/day.
Taking into account that this run is a factor of 14 short in the number
of objects, we would expect a detection rate of approximately
1 object/day
using the full population Eq.\ \ref{neapopulation} with
$V_{\rm lim}=15$. 

Run 2 utilizes the full NEA population Eq.\ \ref{neapopulation} 
with $V_{\rm lim}=15$. From the original approx.\ $1.4\times 10^9$ 
objects, around 10 million, or 0.7\% , were left after the MOID 
sorting.
We find that 16 objects could be seen over 50 days, amounting
to a detection rate of 0.3 objects/day. This number is the same as
for the test population, Run 1. However for Run 1 three objects have
$D\leq 10$m ($H=27.68$, $p_V=0.17$) whereas
of the 16 objects in Run 2, 8 have $D\leq 10$m, and 
3 have $D=1$m. 

These results
we can compare to the observations by Brown and 
collaborators \cite{brown2002}. 
According to their
work, approx.\ 30 objects/yr collides with the Earth, 
with diameters $D\ge 1$m.
To "collide with the Earth" would in the case of Bering 
mean that the object
is detected  ($V\leq V_{\rm lim}$) and encounters Bering within
a distance equal to the Earth radius. On the other hand, an object
within that distance is not necessarily seen by Bering, as the
magnitude of the object depends on the phase angle and the distance
to the Sun and Bering (an object does not need to have $V<V_{\rm lim}$ in
order to collide with the Earth). Hence, in the best 
case, for an object with
$D=1$m
at 1 AU distance to the Sun and at $\alpha=0$, the detection
radius of Bering would be 43260 km at $V=15$ and 6856 km at $V=11$.
For $V=11$, we find three objects in that range, at $V=11.42, V=11.70,
V=11.74$ during the 50 days time span. This amounts to 22 objects/year,
comparable to the 30 objects/yr by Brown et al. If we strictly
look at the diameter,
for $D\leq 4$m, the Earth is hit one time per
year according to Brown et al.
Hence the Earth is
hit by 29 objects/year with $1 {\rm m} \le D \le 4 {\rm m}$. In 
that size range,
we have from the simulation that
6 obj/50d were detected, equal to 44 obj/yr.
With the differences between "colliding with the Earth" and "observing
an object", and the short integration time span in mind (Brown et al.\
is based on 8.5 yr of observations),
we are quite convinced that our
simulations are consistent with the observations. 

For objects smaller than 1m it has been suggested
\cite{bland2003} that the Earth is hit by one object
per day with $D\approx 15 $cm. For Bering, such an object would
have $V\leq 15$ within a distance of 6548 km. 

An even more detailed
integration of the smallest object of
Run 2 reveals that the object would be visible for 43 minutes,
with a close approach distance of 19500 km. The angular velocities
of the smallest of the objects were in the range 100-300 "/sec.
In comparison, a 15 cm object would be moving with 619 "/sec
at the maximum detection distance.

From these considerations, we find that a detection limit of
$V_{\rm lim} = 15$ would provide high feasibility of the Bering 
mission.
Concerning the large angular velocities, there may however
be a conflict between the detection limit and the fastest moving
objects, due to trailing losses. The
capabilities of the ASC to handle fast moving
objects at the suggested limiting detection magnitude remains
to be studied.

As it was mentioned previously, there are strong indications
that the albedo of the smallest asteroids is relatively large.
To check the influence of the albedo on the detection rate,
we for Run 3
simply changed the slope parameter of the MOID-sorted
objects from Run 2, for a direct comparison. 
$G$, the
slope parameter, was changed from 0.15 to 0.32, corresponding
to a change in albedo from approx.\ 0.17 to 0.25. In this run,
23 objects were detected, 7 more than in Run 2, corresponding
to a detection rate of 0.46 objects/day.
14 objects were smaller than 
$H>28$ ($D<7\textrm{m}, p_V=0.25)$. In Run 2, 8 objects had $H>28$,
hence 6 of the 7 additional objects belongs to the small
end of the size range. This indicates that the albedo
may influence the detection rate by a factor of two. This,
on the other hand, may mean that the detection rate measured
by the Bering mission will not only provide a constraint on
the size/absolute magnitude distribution, but also on the
albedo. Particularly in the case
where the Bering observations are consistent with the
ground based observations, the albedo can be
determined directly from the count statistics.
This question should
be address by a more detailed analysis, involving integrations
over longer time spans.

\section{Results for the main belt}

With the value of $V_{\rm lim}=15$ found suitable for the NEA population,
we continued to examine the capabilities of the mission with such
a limiting magnitude in the asteroid main belt. Initially, Bering
was kept at a circular orbit in the ecliptic, with a semi-major
axis of 2.5 AU. This places Bering in the inner part of the main
belt.

A first run, Run 4, was made to test the behavior of the simulations.
Of the 400 million objects, 11 million or 2.7\% survived the
MOID. The integration was made over 356 days, with a resolution
of 1 day. 647 objects were seen, i.e.\ a detection rate
of 1.8 object/day. Of these, 204 were smaller than 10 km ($H>12.94$), 
on the
other hand, none of the smallest objects were seen, the smallest
having $H=27.00$ ($D=15\textrm{m}$, $p_V=0.17)$. Hence
only
a fraction of the largest objects could be seen over one year.
It remains to be examined how many of the largest objects that
can be observed during the mission life time.

The smallest objects were presumably not observable due to the 
transient brightness behavior upon a close encounter, and were
thus not found as the time between each magnitude calculation
($t_{\rm dump}$) is too large.
Hence we made a second run, Run 5, with a time span of 50 days
and with a resolution of 0.02 days. This time, 352 objects could be
seen, or 7 objects/day. Of these, 55 have $H>12.94$, so at least 
on object smaller than 10 km could be seen per day. 
The smallest object found had $H=27.00$, and
was visible for 6 hours, so none of the $H=27.94$
objects could be seen. As mentioned, this may be due to the 
underrepresentation in the number of these objects, in combination
with the short integration time span. It should be examined
whether any of the $H=27.94$ objects can be seen by extending
the integration time span. 
The tangential velocities of the encountered objects were typically
in the range of 5-20 "/sec.

\section{Conclusion}

We have analyzed the scientific feasibility of the Bering mission,
when applied to the Near-Earth Asteroids and the asteroid main belt.
We have found that with a limiting magnitude of the ASC of
approx.\ $V_{\rm lim}=15$,
the scientific feasibility can be sustained.
This would allow the detection of around one object per day, 
in both the NEA and main belt populations. This initial detection
by the ASC
will allow the science telescope to be pointed toward the
object, and detailed observations can be initiated, both in terms
of physical characterisation, but also in terms of following the
object toward faint magnitudes.

For the main belt population, we find that there is room
for lowering $V_{\rm lim}$ if e.g.\ required by the 
smaller power budget.

Some issues however still remains open, in particular the capabilities
of the ASC to follow very fast moving objects at faint magnitudes.
It also remains to be examined how Bering will
perform when placed in the proposed
eccentric orbit
\cite{hansen2003}
ranging from 0.7 AU at perihelion to 3.5 AU at aphelion.
In addition to the already outlined scientific
objectives \cite{andersen2003}, we also find that the albedo of the
object populations has an
influence on the detection rate, and the question of whether 
the albedo can be derived directly from detections among the NEA
population needs a further close analysis, as this is critical
for the physical characterisation, including
the object size.

\section*{Acknowledgments}
Computing facilities were provided by the Aarhus
branch of the Danish Center for Scientific Computing (CSC-AA).
This work was supported by the Danish Natural Science Research 
Council through a grant from the Center for Ground-Based 
Observational Astronomy
(IJAF).

\end{document}